\documentclass[dvipsnames,format=sigconf]{acmart}

\AtBeginDocument{%
  }

\copyrightyear{2024}
\acmYear{2024}
\setcopyright{rightsretained}
\acmConference[GECCO '24]{Genetic and Evolutionary Computation
Conference}{July 14--18, 2024}{Melbourne, VIC, Australia}
\acmBooktitle{Genetic and Evolutionary Computation Conference (GECCO '24),
July 14--18, 2024, Melbourne, VIC, Australia}
\acmDOI{10.1145/3638529.3654033}
\acmISBN{979-8-4007-0494-9/24/07}

\usepackage{dblfloatfix}

\usepackage{graphicx}
\usepackage{booktabs}
\usepackage[compact]{titlesec}

\usepackage[export]{adjustbox}
\usepackage{multirow}
\usepackage{etoolbox}

\usepackage{enumitem}
\usepackage{arydshln}

\begin{document}

\title{Socialz: Multi-Feature Social Fuzz Testing}

\author{Francisco Zanartu}
\email{francisco.zanartu@adelaide.edu.au,ctreude@smu.edu.sg,markus.wagner@monash.edu}
\author{Christoph Treude}

\author{Markus Wagner}

\renewcommand{\shortauthors}{Zanartu et al.}

\begin{abstract}
Online social networks have become an integral aspect of our daily lives and play a crucial role in shaping our relationships with others. However, bugs and glitches, even minor ones, can cause anything from frustrating problems to serious data leaks that can have far-reaching impacts on millions of users.

To mitigate these risks, fuzz testing, a method of testing with randomised inputs, can provide increased confidence in the correct functioning of a social network. However, implementing traditional fuzz testing methods can be prohibitively difficult or impractical for programmers outside of the social network's development team. 

To tackle this challenge, we present Socialz, a novel approach to social fuzz testing that 
(1) characterises real users of a social network, 
(2) diversifies their interaction using evolutionary computation across multiple, non-trivial features, and 
(3) collects performance data as these interactions are executed. 
With Socialz, we aim to put social testing tools in everybody's hands, thereby improving the reliability and security of social networks used worldwide.

In our study, we came across (1) one known limitation of the current GitLab~CE and (2) 6,907 errors, of which 40.16\% are beyond our debugging skills. 
\end{abstract}




\maketitle

\sloppy

\section{Introduction}

Online social networks (OSNs) are an integral part of modern society, influencing a wide range of aspects of our daily lives. The vast quantity of personal information shared on these platforms makes them a treasure trove for companies seeking to reach out to potential customers and for individuals looking to grow their social circle or entertain themselves. However, like any software, OSNs are prone to bugs and technical issues; consequences range from poor user experience~\cite{brodkin2022socialapp,day2015socialbugs} to massive data breaches affecting billions of individuals~\cite{morahan2022dataleaks}.

The recent rise of social bugs in software systems has prompted the need for social testing~\cite{ahlgren2020wes}. However, for the research community, social testing poses several challenges. 
First and foremost, obtaining data from online social networks (OSNs) can be time-consuming, resource-intensive, and requires specialised expertise, which may not be accessible to non-specialists~\cite{https://doi.org/10.48550/arxiv.1612.04666}. For example, privacy policies and community guidelines may restrict access to data, making it unavailable to researchers; Furthermore, descriptions of data extraction methods are often omitted in many studies~\cite{info:doi/10.2196/13544}. 
Second, researchers may be limited in their ability to conduct experiments on OSN systems if they are built using proprietary software platforms~\cite{ahlgren2020wes}. 
Finally, the sheer size and complexity of OSNs can result in significant operational costs, which can impede researchers' ability to conduct large-scale experiments, limiting the scope and depth of their research.

To overcome these challenges, technology companies develop tools such as Web-Enabled Simulation (WES, by Facebook/Meta)~\cite{ahlgren2020wes}, allowing developers to test code updates and new features in a simulated environment, without risking real user data. 
Social testing, which involves simulating interactions among a large community of users, can be used to uncover faults in online social networks. 
However, these tools are not available to the public and may have limitations in simulating the full spectrum of user behaviours. 

To address these limitations, we introduce Socialz, an approach for social fuzz testing, which makes the following key contributions:

\begin{enumerate}[topsep=2mm]
    \item characterisation of users of a real social network, 
    \item evolutionary diversification of community interaction with respect to multiple, non-trivial features, 
and 
    \item a workflow for executing interactions and collecting performance data.
\end{enumerate}

Socialz aims to advance the field of social testing through diversity-based user behaviour: it evolves diverse sets of virtual users that are distributed across a non-trivial feature space, which in turn enables us to cover a wider range of behaviours compared to real users, and thus increases the likelihood of uncovering bugs that may not be detected by a set of similar and biased virtual users. 

\section{Related Work}

Software testing has evolved into a vast field that includes many different methods and techniques for assessing the performance, usability, and other attributes of software systems~\cite{7814898}. Testing is performed at various levels of abstraction, ranging from unit testing to system testing.

Recently, the concept of social testing has emerged as a new level of abstraction, potentially positioned above system testing~\cite{ahlgren2020wes}. This is due to the recognition that social bugs can arise through community interactions and may not be uncovered by traditional testing that focuses solely on single user journeys~\cite{harman2018journeys}. 

Next, we briefly survey prior work in the testing of social networks, the application of evolutionary methods in fuzz testing, and diversity optimisation.

\subsection{Testing of Social Networks}

Search-Based Software Testing (SBST) is a technique that leverages optimization search algorithms to solve software testing problems~\cite{5954405}. This approach is widely used in both industrial and academic sectors, including at Facebook, where it is employed to test the behaviour of both the system and its users~\cite{10.1007/978-3-319-99241-9_1}.

In social networks, testing goes beyond simply assessing the behaviour of the system and involves evaluating the interactions between users facilitated by the platform. To that end, Web-Enabled Simulation (WES) simulates the behaviours of a community of users on a software platform using a set of bots~\cite{ahlgren2020wes}. Traditional tests, on the other hand, involve executing a predetermined series of input steps. WES is run in-vivo but in a shadow copy of the system, hence as a separate ``digital twin'', which allows for testing without risk to real user data~\cite{ahlgren2021facebook}. Both ``digital twins'' and WES have widespread industrial applications, not only in OSNs but also in robotics, manufacturing, healthcare, and  transport~\cite{jiang2021industrial,8972429}.

In addition to testing, there are also methods for formally verifying the correctness of social network software and systems. Formal verification of social network protocols and algorithms ensures security and reliability by ensuring access to content is subject to both user-specified and system-specified policies~\cite{cheng2012user,kafali2014detecting}. Model checking can be applied at various levels of abstraction, from high-level network properties to the implementation of individual components~\cite{pardo2017model}. This approach can be used in combination with other testing techniques, such as simulation or testing with software-controlled bots, to provide a more comprehensive evaluation of a social network's behaviour and performance~\cite{pedersen2021social}.

\subsection{Evolutionary Fuzzing}

We adopt definitions from~\citet{https://doi.org/10.48550/arxiv.1812.00140}, who define fuzzing as a software testing methodology  that involves injecting unexpected or randomised input data into a program-under-test to uncover defects or bugs. A specific application of fuzzing, called fuzz testing, evaluates the security policy violations of the PUT. The tool used to perform fuzz testing is known as a fuzzer.

The inputs used in fuzzing can be selected either randomly, with each element having an equal chance of being chosen, or through guided methods such as syntactic or semantic models~\cite{MARIANI2015bbtesting}. 
Evolutionary fuzz testing is a variant of fuzz testing that leverages evolutionary algorithms to optimise the input data. Research has shown that it is effective in discovering a wide range of vulnerabilities, including those that are difficult to detect with traditional testing methods. For example, \citet{li2019v} demonstrate V-Fuzz, a vulnerability-oriented evolutionary fuzzing framework that combines vulnerability prediction with evolutionary fuzzing to 
reach potentially vulnerable code. They test various open-source Linux applications and fuzzing benchmarks. 
For Android applications, \citet{Cotroneo_2019} introduce a coverage-guided fuzzing platform that demonstrated to be more efficient than blind fuzzing. 
Similarly, \citet{9847081} employ a constrained neural network evolutionary search method to optimise the testing process and efficiently search for traffic violations.

Although fuzzing is widely used, to the best of our knowledge, no research has been conducted to fuzz test a social network system. 

\subsection{Diversity Optimisation}\label{sec:diversityopt}

When it comes to social fuzz testing, the actions a user takes need to be determined, such as ``follow a particular person''. In such a case, defining a user that achieves exactly this goal is straightforward, as a developer can easily translate the desired outcome into a specific interaction. 
However, creating a user with a more complex behaviour, such as ``a virtual user should be highly active but not very central to everything that is going on'', is not as simple. Mapping the desired interaction between the virtual user and its environment is challenging due to the intricate interplay, even though the activity and centrality calculations may not be black boxes by definition. 
This issue becomes even more complex when designing a group of virtual users within a social network that can interact with each other in various ways.

We propose a practical solution to addressing such issues by treating the user features as black box functions and utilising heuristic approaches like novelty search~\cite{DBLP:conf/gecco/RisiVHS09} or 
evolutionary diversity optimisation~\cite{DBLP:conf/gecco/UlrichT11}; 
this has been made possible only recently by algorithmic advancements in diversity optimisation with multiple features. 

\section{The Methodology of Socialz: Overview}\label{sec:methodology}


We define social fuzz testing as a method aimed at finding bugs in online social networks by simulating diverse user behaviours and interactions. By systematically diversifying these interactions, social fuzz testing aims to reveal bugs that traditional social testing might overlook.

Socialz instantiates this via three-stage approach, each of which has a distinct methodology: To establish a realistic baseline for user interactions, in \textit{Stage 1/3: Characterisation of User Behaviour}, we detail the process of obtaining and analysing data from a real OSN to understand the behaviour of users on the network.
To simulate and enhance unexpected user behaviours that may challenge the robustness of OSNs, \textit{Stage 2/3: Evolutionary Diversification of Community Interaction} then improves the diversity of user behaviour in the network.
Finally, to validate the resilience of OSNs against diversified user behaviours, \textit{Stage 3/3: Execution} involves evaluating the evolved network.


\paragraph{Target platform.} \label{sec:gitlab} 
To demonstrate Socialz, we need to choose a server as the PUT. The server needs to  meet certain requirements to ensure that the case study can be conducted, 
such as being freely available, providing an API to impersonate users, and providing the ability to gather system performance data.

\begin{figure}
\centering\vspace{-2mm}%
\includegraphics[width=0.9\linewidth,trim=30 0 0 10,clip]{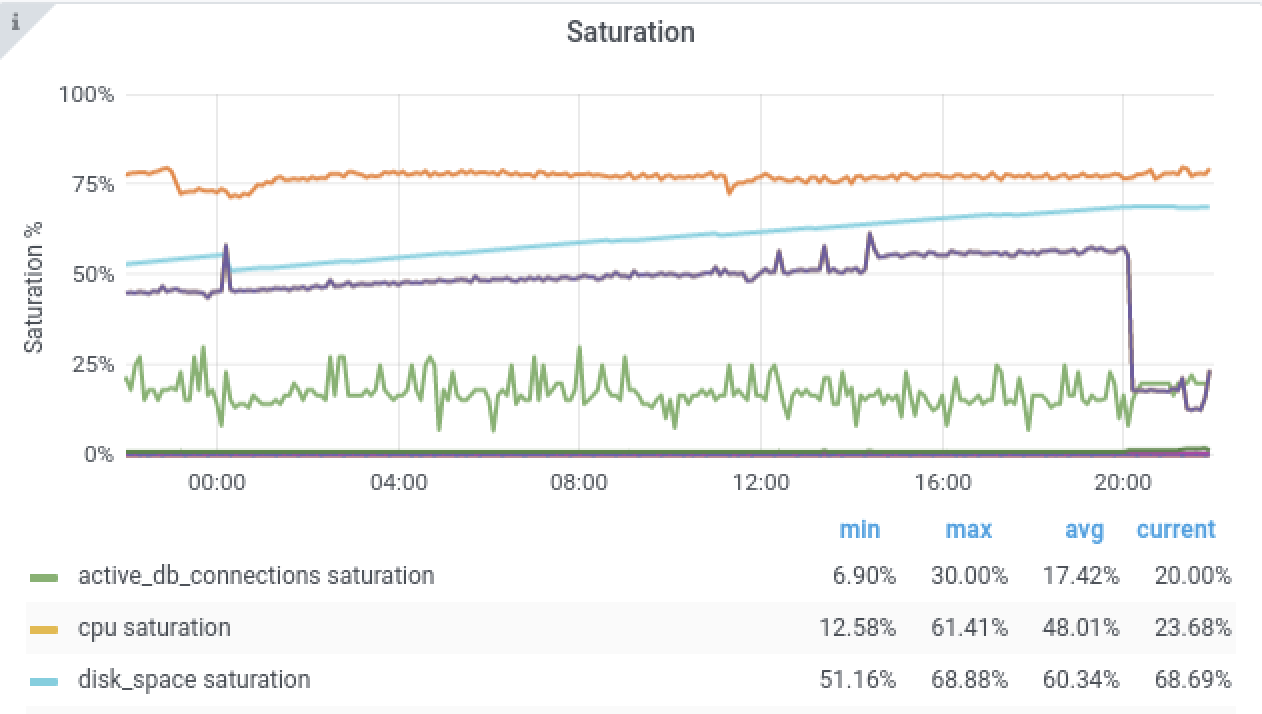}\vspace{-2mm}
\caption{Grafana dashboard example, showing statistics of our GitLab~CE server.}
\label{fig:grafana}\vspace{-2mm}
\end{figure}

GitLab Community Edition (GitLab~CE) is a good fit for these requirements. 
GitLab~CE is the free and open source edition of GitLab, which is a platform that has over 30 million registered users~\cite{gitlabusers}. 
Despite being free, GitLab~CE provides a comprehensive set of performance metrics that are continually stored on an internal Prometheus time-series database. This database can be scrapped with a comprehensive set of pre-defined Grafana dashboards~\cite{gitlabdocumentation}, providing a wide range of performance metrics for our purposes (see Figure~\ref{fig:grafana}).

To easily reset the system to well-defined states, we run GitLab~CE 16.6 (November 2023) in virtual machines 
with Ubuntu 22.04 LTS. For the evolutions, we utilised the CPU optimised c3.xlarge 
instances (VCPUs: 16, memory: 32 GB), and for loading the data smaller CPU optimised c3.medium 
(VCPUs: 4, memory: 8 GB).

\section{Characterisation of User Behaviour (Stage 1/3)}

As the starting point in our case study, we populate our GitLab server with real-world data from GitHub. We use GitHub as our data source because it has a large community of over 100 million developers~\cite{dohmke2023million}. 
There are also several Git server alternatives, such as Gitea and Gogs\footnote{\url{https://gitea.io}, \url{https://gogs.io}.}, that can be self-hosted. 
Gogs is a lightweight self-hosted Git service that can run with minimal computational requirements. Gitea is also a lightweight application, originally forked from Gogs, that offers more features such as code hosting, code review, CI/CD, team collaboration and package registry. However, at the moment of our project, neither of both options included a complete set of metrics to monitor the application performance, and the development of such metrics would have required a significant effort beyond the scope of our project. In contrast, GitLab stores its metrics in a Prometheus time-series database that can be scrapped or accessed with any HTTP client. Moreover, these metrics can be easily monitored online with a set of pre-configured dashboards offered by GitLab.

The next point is as critical as the existence of a server that we can test: projects exist that provide access to user interaction data that is similar to that found in online social networks (OSNs). These interactions include creating and annotating content, creating networks (e.g. starring a repository, linking comments), as well as ``malicious'' interactions (e.g. intentionally or unintentionally submitting bugs, spamming, or violating privacy). One such project that provides data on interactions is GH~Archive\footnote{\url{https://www.gharchive.org}.}. It is a public dataset available on Google Big Query that has been recording the public GitHub timeline since 2011, and it makes this data easily accessible through SQL-like queries of JSON-encoded events as reported by the GitHub API.

To narrow down the immense amount of data available in GH~Archive\footnote{As of June 2023, the GH~Archive data stored on Google BigQuery totals more than 19.64 terabytes.}, we choose a sub-community within GitHub, and after considering various sub-communities based on their size, we find that COBOL's is small enough to enable us to conduct a thorough analysis of the data. This is an intentional choice, because we target a ``complete'' subset of the data, i.e. not a random sample of nodes or edges from GH~Archive that are not interconnected. We model the data as a graph, with users and repositories as nodes and GitHub events as edges. In this model, repositories can be thought of as groups on an online social network (OSN) where users share and contribute content. 

Next, we explicitly establish similarities between GitHub events to activities as they can be found on other networks. In particular, we focus on GitHub events that are akin to content creation, content annotation, and network creation: 
\begin{enumerate}[topsep=2mm]
\item \textit{WatchEvents} and \textit{ForkEvents} can be likened to liking a public profile page.
\item \textit{PushEvents} can be thought of as being invited to a group with permission to publish some content.
\item \textit{PullRequests} can be thought of as requesting permission to publish something to a group.
\item \textit{FollowEvents} represent establishing a connection or friendship with another user. Unfortunately, as of December 2013 FollowEvents have stopped being recorded in GH~Archive, we need to create a workaround where connections between users are based on their similarity. To ensure consistency in our analysis, we disregard the existing FollowEvents in the GH~Archive data and instead utilise only our own approach (see Section~\ref{sec:solrep}).
\end{enumerate}

Finally, to ensure that our dataset is as complete as possible, we further adjust its size by filtering it to only include events from the years 2011 to 2016 (see Table~\ref{tab:original}). This decision allows us to compile a relatively complete dataset (i.e. starting from the beginning of GH~Archive's records and going up to a particular date), rather than having more recent but incomplete data (e.g. like considering the last six years until today) in which possibly all relevant events would have occurred before the starting date of the snapshot. 


\begin{table}
\centering\vspace{-2mm}
\caption{Original dataset: 1,523 users created a total of 6,742 events involving 156 repositories and forks (2011--2016).}
\label{tab:original}\vspace{-2mm}
\begin{tabular}{cc}
  \toprule
  \textbf{Event type} & \textbf{Number of events} \\
  \midrule
  PushEvent & 4234 \\
  WatchEvent & 1206 \\
  PullRequestEvent & 852 \\
  ForkEvent & 450 \\
  \bottomrule
\end{tabular}\vspace{-2mm}
\end{table}

\section{Evolutionary Diversification of Community Interaction (Stage 2/3)}

In this section, we define the components of our evolutionary approach and how they are used to diversify a set of virtual users, which are less biased than their real-world counterparts. This process has the potential to reveal anomalies or unexpected behaviours that would otherwise be difficult to detect in sets of bots that are similar and biased.

\subsection{Features of Community Interaction}

To characterise users, we investigate three features that we consider to be non-trivial in the sense introduced in Section~\ref{sec:diversityopt}: given these features, a developer may struggle to manually design a set of virtual users that exhibit a spread of desired community interactions. 
Our chosen features allow us to characterise how active a user is, what is its relative importance and what kind of interactions the user is involved in: 
First, the graph \textit{degree of centrality} measure of the nodes quantifies how active a user is in the network, i.e. how many events are submitted. This measure is often used as a notion of popularity in social networks~\cite{https://doi.org/10.48550/arxiv.2011.01627}, as nodes with a large number of relationships are considered more powerful and central, but has a limitation in that it only takes into account local knowledge of the network topology. Hence, we introduce an additional centrality metric to supplement its analysis next. 

Second, to assess the relative importance of a user on the network, we utilise the \textit{PageRank} algorithm~\cite{page1999pagerank}: it is fast to compute, well suited for a directed network such as ours and has been proven to be effective in characterising users~\cite{9420317}.

Third, to characterise the types of actions a user performs --- for example, a user may submit only PushEvents, or only ForkEvents and PullRequestsEvents --- we represent each combination as a binary vector and then consider the corresponding decimal value as that user's \textit{event type}.\footnote{Because the number of users is much larger than the number of possible combinations, and because we aim for diversity, we conjecture that variations to this mapping procedure only have minor effects on the overall outcomes.} We consider 15 combinations, as we have four event types that are not FollowEvents, and the combination of ``user does not interact at all'' is not allowed.

It is critical to emphasise that these three metrics are features, not objectives: no user is ``better'' or ``worse'' than another one, neither in a single-objective sense, nor in a multi-objective sense.

\subsection{Solution Evaluation}\label{sec:solrep}

In our evolutionary setup, each individual is an interaction graph that represents how virtual users interact in a social network. 
In particular, each individual is an edge list that contains all the necessary information of our graph: the source node, the target node, and the type of event. 
To evaluate our graphs, we have defined the following six steps. 

First, we transform the edge list into an two-dimensional adjacency matrix. This adjacency matrix has four areas that reflect the interactions repo-repo, repo-user, user-repo, and user-user. 

Second, we record the interactions between users and repositories in the repo-user and user-repo areas by summing the number of events that occur between each repository and user. Each event has a weight of one, regardless of its type. 
At present, as we do not further differentiate, the repo-user and the user-repo areas are mirrored versions of each other. 

Third, as there are no interactions between repositories in our approach, the repo-repo area is always filled with zeros.

Fourth, we use the user-user area to store FollowEvents. It is initially empty, and we fill it by evaluating the cosine similarity of the users based on their interactions with repositories (areas user-repo, repo-user). 
If the cosine similarity is greater than zero for a pair of two users, then we set the respective entry to 1. 
The diagonal is set to 0 as users cannot follow themselves. The resulting matrix for the Original dataset looks like this (users 658 and 659 are chosen to show non-zero data, as the matrix is sparsely populated): 
\vspace{-3mm}
\begin{center}
{\scriptsize{\setlength{\tabcolsep}{1.5mm}
\begin{tabular}{c|ccc:cccccc|}
      & repo$_1$ & $\cdots$ & \multicolumn{1}{c}{repo$_{156}$} & user$_1$ & $\cdots$ & user$_{658}$ & user$_{659}$ & $\cdots$ &   \multicolumn{1}{c}{user$_{1523}$} \\ \hline
repo$_1$ & 0         &     & 0     & 0    & & 0&4     &     & 0     \\
$\vdots$   &           &     &       &       &  &  &    &     &       \\
repo$_{156}$ & 0        &     & 0     & 0   &  & 0 & 0    &     & 0     \\ \cdashline{2-10}
user$_1$ & 0        &     & 0     & 0    & & 1 & 0    &     & 0     \\
$\vdots$ &&&&&&&&&\\
user$_{658}$ & 0        &     & 0     & 1    & & 0 &0     &     & 0     \\
user$_{659}$ & 4        &     & 0     & 0    & & 0 & 0 &        & 0     \\
$\vdots$   &    &       &     &       &       &       &  &    &       \\ 
user$_{1523}$     & 0     &     & 0     & 0   &  & 0 & 0    &     & 0 \\ \cline{2-10}
\end{tabular}}}
\end{center}

Fifth, with the FollowEvents incorporated into our edge list, we calculate the PageRank score and degree of centrality of each node. For the event type feature, we filter our data to only include user nodes and map each user node to the combination of events they were involved in.

Finally, we calculate the star-discrepancy score for the interaction graph. 

Our rationale for the decision to create FollowEvents between two users based on the simple criterion that the cosine similarity across all events (for these two users) is greater than zero is three-fold and mostly based on practical considerations: (1) we make the assumption that users who create similar events may be likely to follow each other, (2) the approach is deterministic and thus saves memory (at the cost of computation time), and (3) it reduces the search space by allowing us to generate community interactions. 

\subsection{Evolutionary Algorithm}

We employ a diversity optimisation approach using the star-discrepancy measure, based on~\cite{neumann2018discrepancy}. The star-discrepancy measures the regularity with which points are distributed in a hypercube, and in particular with respect to all axis-parallel boxes $\left[0,\ b\right]$, $b\ \in\left[0,\ 1\right]^d$ that are anchored in the origin. Hence, this metric helps us evaluate how evenly the points are distributed in the feature space. In our case, each point  represents a user with its coordinates defined by the three above-described metrics. 
We linearly scale all three metrics into $\left[ 0,1 \right]$.

We use a $\left(1+20\right)$-EA, and in each mutation, we randomly add and delete edges, where the particular action and the particular edge are chosen uniformly at random. 
When deleting edges, we do not allow nodes to be disconnected, in which case we resample. 

To aid the convergence, we utilise a success-based multiplicative update scheme that can lead to faster solution convergence~\cite{doerr2018parameterselection}. This scheme provides a dynamic mutation rate for the EA based on the performance of the offspring. If an iteration is successful, meaning an offspring is not worse than the current solution, the mutation rate is increased by a constant factor $A=2$. If the offspring is not better, the mutation rate is decreased by a constant factor $b=0.5$. The initial per-edge mutation rate is $1/n$, where $n$ is the number of edges, which is the total number of events that are not FollowEvents.

In summary, our evolutionary approach to diversified community interaction works as follows. We pass an interaction graph as an edge list to our evolutionary algorithm, mutate the edges as described, compute the user’s similarity to add FollowEvents and then create our graph to compute PageRank and centrality for each user and map the users to the combination of events they were involved with. With these three features, we compute a graph's star-discrepancy score and, by means of our evolutionary algorithm, iteratively keep improving. 

\subsection{Diversified Community Interaction}

We start by using the original edge list of 6,742 events to calculate the similarity between the 1,523 users, resulting in an edge list of 397,224 events with a star-discrepancy score of 0.305 in the three-dimensional feature space. Then, we use this as the initial definition of the community interaction on our server, and apply the previously described evolutionary approach. 
We perform 30 independent evolutions, each for 10,000 generations, with each run taking approximately 30 hours. 
This results in interaction graphs with star discrepancies between 0.01107 and 0.01166. 

\begin{figure}
\centering\vspace{-2mm}%
\includegraphics[width=0.99\linewidth,trim=0 0 0 40,clip]{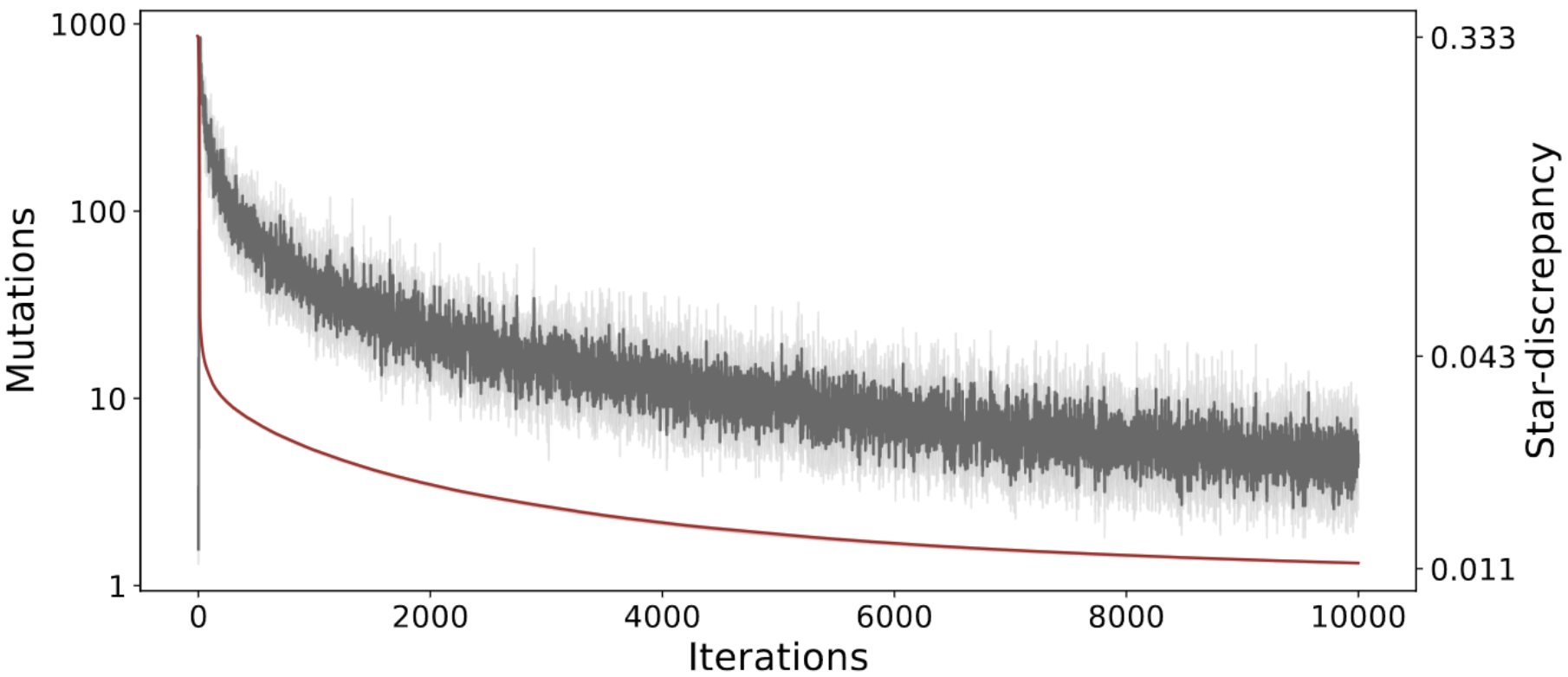}
\vspace{-3mm}
\caption{Evolution of the interaction graph. 
For 30 independent runs, red shows the average discrepancy (of user behaviour), and grey shows the number of mutations (light grey: min-max range; dark grey: 95\% confidence interval).}
\label{fig:convergence}\vspace{-2mm}
\end{figure}

Figure~\ref{fig:convergence} shows the evolution over time. 
The self-adaptive parameter control appears to work well: (1) the number of mutations increases very quickly to several hundreds resulting in a quick improvement in diversity, (2) the number of mutations then gradually decreases to tens of mutations from 6,000 generations on, while still resulting in further improvements of the diversity.

\section{Execution (Stage 3/3)}\label{sec:eval}

In this section, we present our approach for executing and evaluating community interaction. 

\subsection{Benchmarking the Evolutionary Approach}

A natural question is how to compare the Original and Evolved edge lists, because they are of notably different sizes? 
Examplarily, we answer this using a practical approach. First, of the 30 independent runs from Stage 2/3, we select the interaction graph with the lowest star-discrepancy; its evolved edge list contains 433,054 events with a corresponding star-discrepancy score of 0.01107
for 1,523 users.
Next, to compare the Original and Evolved edge lists, we craft additional datasets using two approaches. The first approach creates a larger version (called ``Simple'') of the original edge list by copying only the existing events until the size of this simple version matched that of the evolved edge list. The second approach generates new connections at random until the edge list reached the same size as the evolved one; the resulting community interaction is called ``Random''. In both approaches, we ensure that the number of FollowEvents is the same, as these events are considerably more numerous than other types of events. 
By reducing the potential impact of differences in size on the validity of the results, these approaches allow for the creation of comparable versions of the Original and Evolved interactions.

Table~\ref{table:alldatasets} shows a first comparison of the four interaction graphs: the Original one that is directly based on GitHub data, its Simple (but larger) version, the Random version, and the Evolved one. 

\setlength{\tabcolsep}{1.5mm}
\begin{table}
\centering\vspace{-2mm}
\caption{Dataset comparison: number of events}\vspace{-2mm}
\label{table:alldatasets}
\begin{tabular}{ccccc}
  \toprule
  \textbf{Event type} & \textbf{Original} & \textbf{Simple} & \textbf{Random} & \textbf{Evolved}\\
  \midrule
  FollowEvent & 390,482 & 420,502 & 420,502 & 420,502 \\
  PushEvent & 4,234  & 7,895 & 5,644 & 4,822\\
  WatchEvent & 1,206  & 2,293 & 2,666 & 2,576\\
  PullRequestEvent & 852  & 1,561 & 2,367 & 2,578\\
  ForkEvent & 450  & 803 & 1,875 & 2,576\\
  \midrule
  \textbf{Total} & \textbf{397,224} & \textbf{433,054} & \textbf{433,054} & \textbf{433,054} \\ 
  \bottomrule
\end{tabular}\vspace{-2mm}
\end{table}


\begin{figure*}
\centering\vspace{-2mm}%
\includegraphics[width=\textwidth,trim=15 0 15 0,clip]{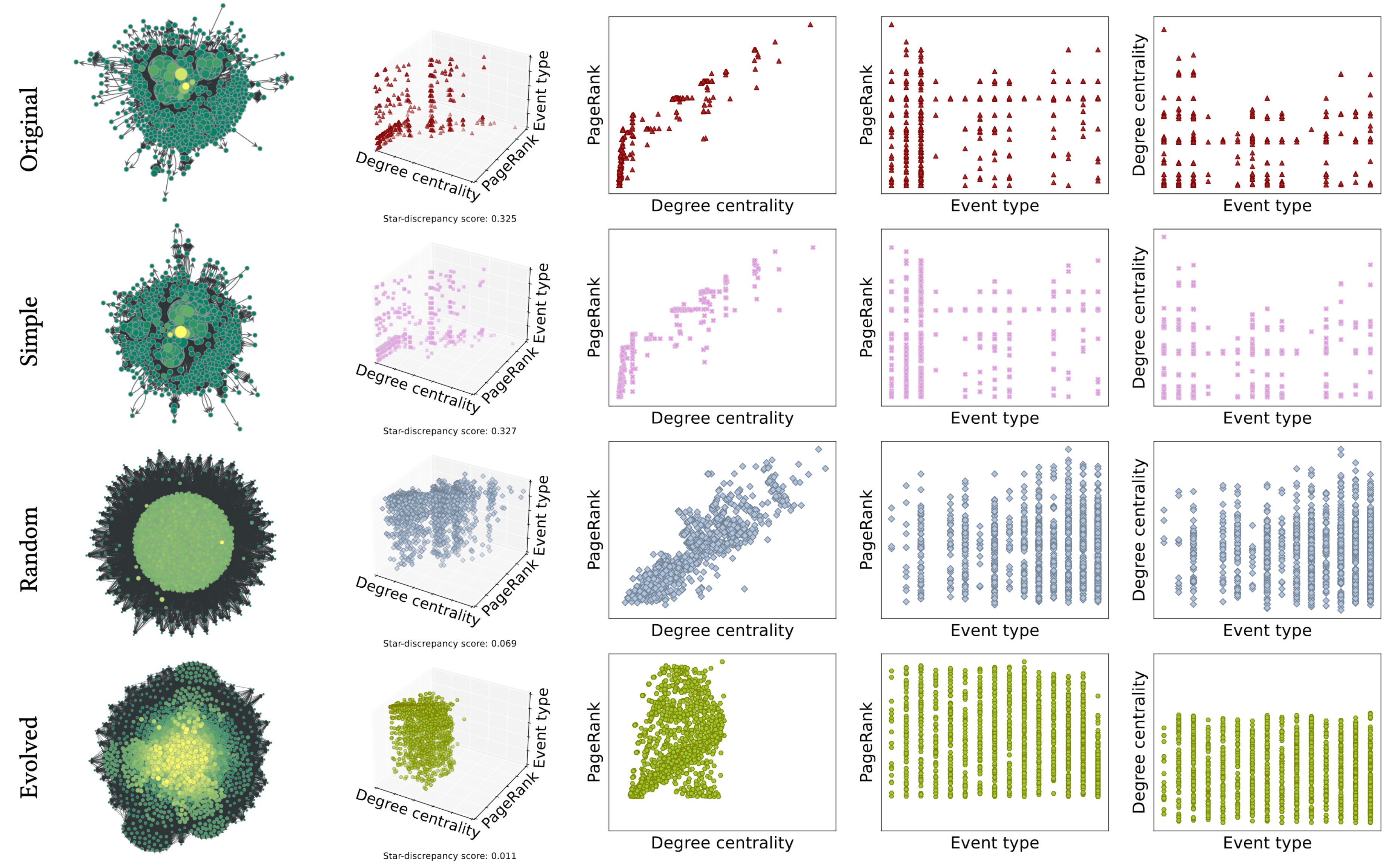}
\vspace{-7mm}%
\caption{Dataset comparison: user interaction based on interaction features. The 2d plots are projections of the 3d plots. The value ranges are always $\left[0,1\right]$ based on the minimum and maximum values across all four interaction graphs.}\vspace{-1mm}%
\label{fig:featurespace}
\end{figure*}

Figure~\ref{fig:featurespace} presents visualisations of the four communities. First, the left-most column shows projections of the graphs into 2d: edges refer to interactions between users and repositories; 
dot size represents the degree centrality of a node with larger dots indicating a higher degree centrality; 
and the colour of each dot represents the PageRank score of the node, where the colour range from green, less important, to yellow, higher score.
As we can see, the Original and the Simple ones are (subjectively) close in structure. The Random one appears fairly homogeneous, and the Evolved one is much more diverse in terms of the distribution of PageRank scores and degree centrality. 

Figure~\ref{fig:featurespace} complements these observations by showing the three features used to calculate the star-discrepancy score for each dataset, i.e. the degree of centrality of each user, their PageRank score, and the combination of events they are involved with. In particular, the discrepancy scores are 0.325 for the Original interaction graph, 0.327 for Simple, 0.069 for Random and 0.011 for Evolved. 

The visualisations show that users in the Original and Simple datasets tend to cluster together and occupy a smaller space, while users in the evolved edge list and the random version of the original edge list appear to be more evenly distributed throughout the space. Interestingly, even the random version achieves a fairly diverse set of interactions, although the event types appear much less covered by the random dataset when compared to the evolved dataset. 
While it may appear that Evolved's diversity in the PageRank and degree centrality subspace is less-than-impressive, in its own subspace ranges, the diversity is very even; Evolved's evolution dropped large degree centrality users, and later (as evolution is diversity focussed), the boundaries would no longer be actively pushed outwards, relative to what can happen during Simple's and Random's construction. 
Despite this limitation, this data suggests that our evolutionary algorithm effectively improves the distribution of users in the feature space.

\subsection{Observing effects of community interactions}

To assess the impact that the different datasets have on the server, we consider the processing of the community interaction as an actual benchmark in itself: as the hundreds of thousands events are processed between the 1523 users on the server, we observe how the system behaves. 
In the following, we outline the workflow used when executing the event and we present our observations.

\subsubsection{Methodology}

We require an elaborate workflow 
as the randomised events create a broad range of situations that need to be dealt with; they would otherwise simply results in a myriad of errors. Essentially, all types of events are first validated by checking GitLab~CE's database to see if the user triggering the event exists. If not, the user is created. The same process is followed for the user and/or repository targeted in the event. The flow then proceeds to the corresponding action for that event.


A complex logic is required for pull request events, where we select or create a branch to submit a pull request. If there is already an open pull request on that branch, we try to merge it. Otherwise, we close the event. If the pull request is closed, we reopen it. To add some realism, 
we use a corpus of words that we extracted from the original dataset, so when we create a commit or a pull request, we add a random text from this corpus, allowing the system to check how many lines of text were added or deleted following Git logic.

On the technical side, we use the previously described setup with GitLab~CE and the virtual machine. 
Our GitLab API wrapper code implements the workflow and is used to load our datasets (Original, Evolved, Simple, and Random). 
During the processing, we collect performance data from the internal Grafana dashboard panels and from the Prometheus database (see Section~\ref{sec:methodology}) included in the GitLab installation. As substantial development effort has gone into developing the evaluation environment, we make the virtual machine images publicly available at \url{https://github.com/fzanart/Socialz/}.

The processing of all community interactions is time-consuming: the execution of the Original/Evolved/Simple/Random datasets on the virtual machines takes between 16 and 42 hours.

\subsubsection{Effects on the system}

\begin{figure}
\centering
\vspace{-1mm}\includegraphics[width=75mm,trim=0 0 0 0,clip]{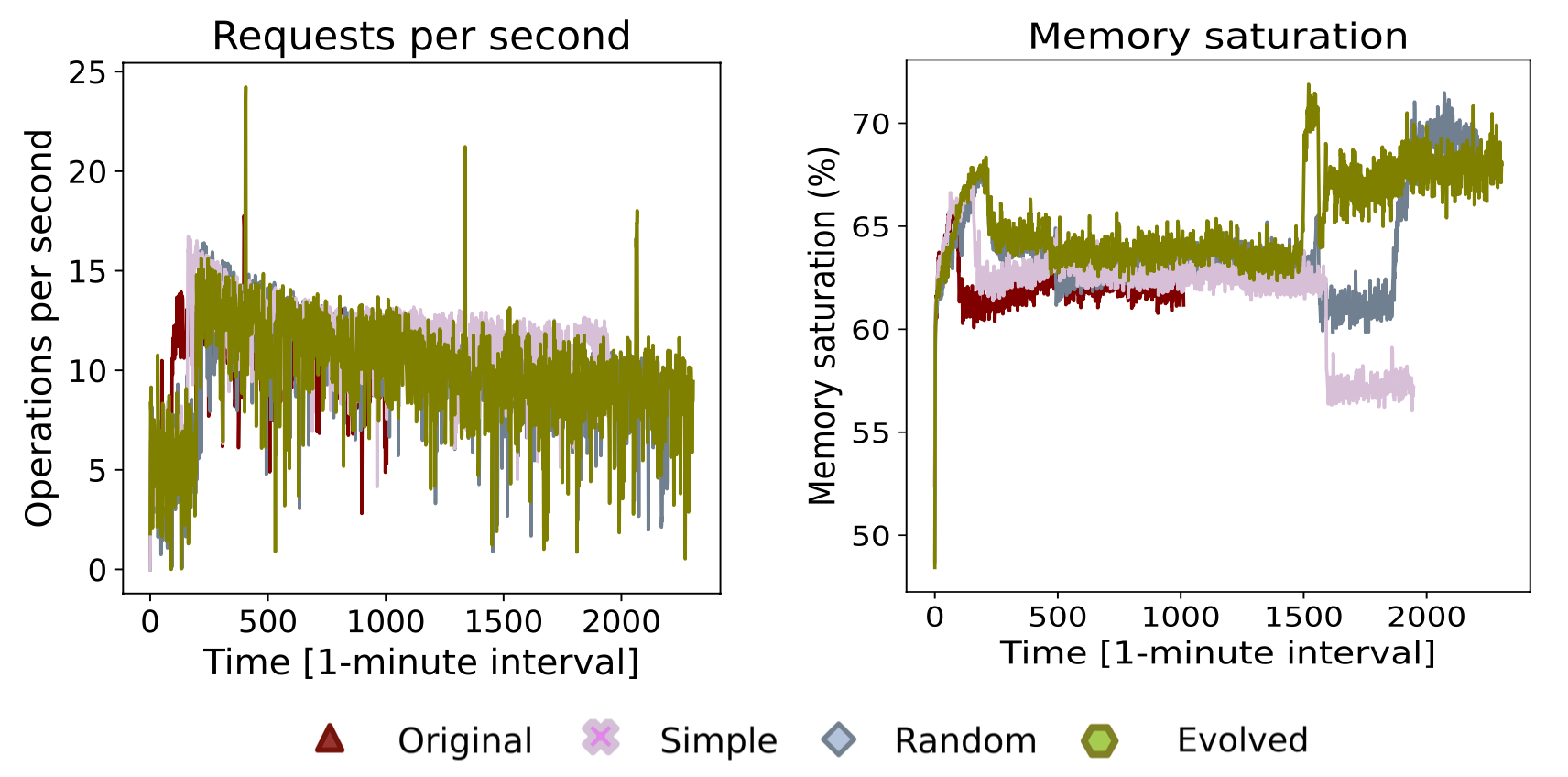}%
\vspace{-3mm}
\caption{Processing of community interactions over time. 
}%
\label{fig:prometheus}\vspace{-2mm}
\end{figure}

First, Figure~\ref{fig:prometheus} shows the requests per second handled (as a proxy for CPU saturation) and memory saturation over time, 
while the interactions are processed. 
As we can see, all four interaction graphs result in different workloads over time. The Original interactions load the fastest as their number is substantially below those of the other three. The Evolved ones appear to have the largest number of spikes in the CPU saturation and the lowest rate of processed requests per second, resulting in the longer overall time needed to load the data.


\begin{figure}
\centering\vspace{-3mm}%
\vspace{-1mm}\includegraphics[width=75mm,trim=0 0 0 0,clip]{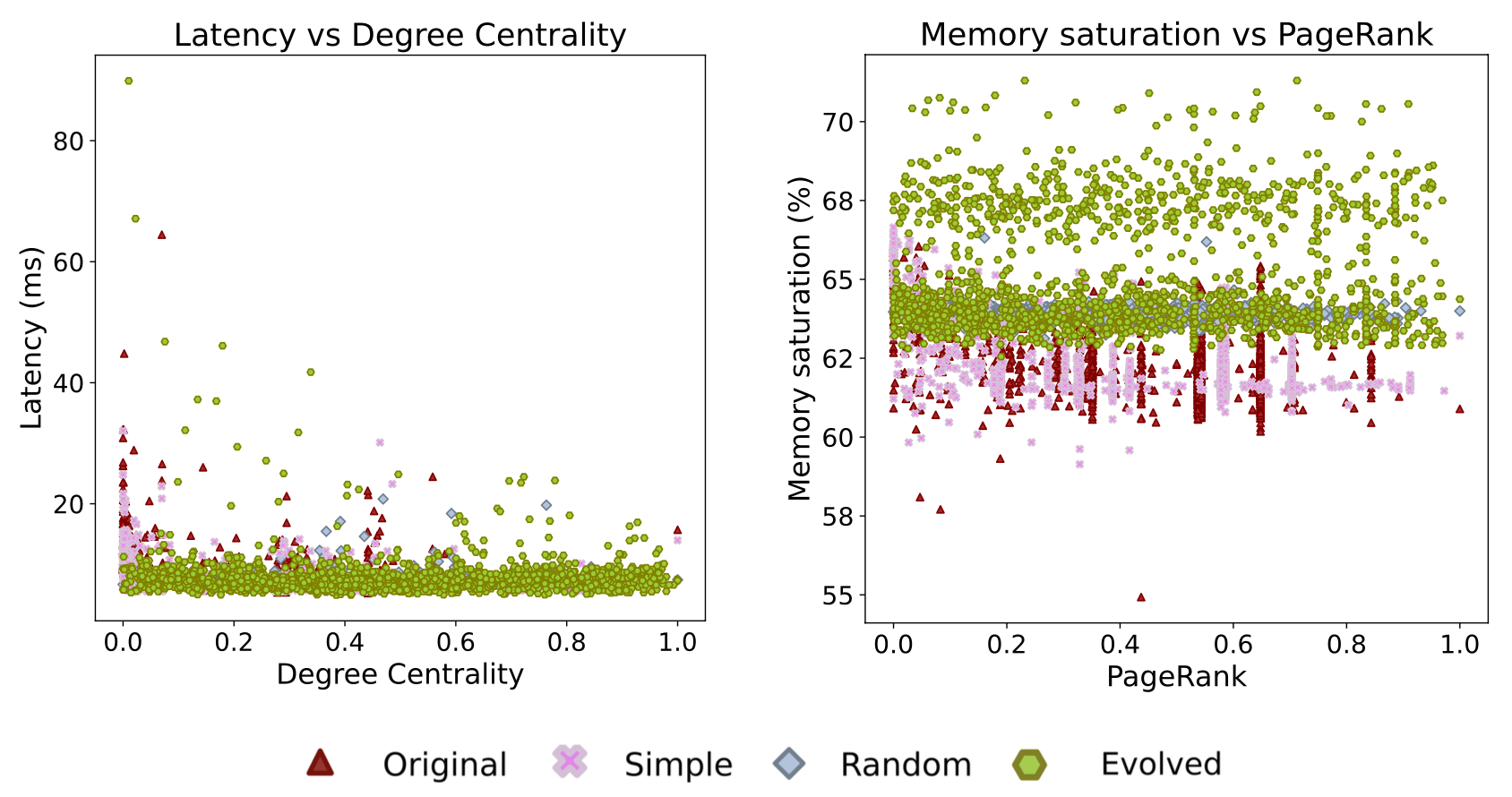}%
\vspace{-3mm}
\caption{Correlation of user features and resource utilisation. Shown are the averages for each of the 1523 users.}
\label{fig:correlation}\vspace{-2mm}
\end{figure}

To gain further insight into user behaviour and its impact on the system, 
Figure~\ref{fig:correlation} also presents a correlation analysis. 
Interestingly, despite acknowledging the fluctuation in performance over time as illustrated in Figure~\ref{fig:prometheus}, our analysis identifies several statistically significant differences in this case. 
In particular, we contrast the resource utilization metrics Latency with the Degree Centrality. We observe medium, negative Spearman correlations ($r \leq -0.36$) for both the Original and Simple datasets (statistically significant). For the Random and Evolved interaction graphs, a very weak, positive correlation ($ 0.05 \leq r \leq 0.07$, statistically significant). 

Similarly, for memory saturation and PageRank, we observe negative correlations ($r \leq -0.22$, statistically significant) for both the Original and Simple datasets. For the Random and Evolved interaction graphs, no statistically significant correlation exists.

\subsubsection{Characterisation of the Performance Indicators}

When evaluating the potential usefulness of performance indicators during the evolution of community interaction in-vivo on a server, we have made three main observations. 

First, we have noticed that neighboring data points in Figure~\ref{fig:prometheus} often show variation due to what seems to be random noise, making it challenging to compare marginal differences that may be impacted by factors outside of our control. In addition, the correlation analysis shows that users who are similar, such as having the same degree centrality, can still have different experiences in terms of server load, making it potentially difficult to develop a surrogate.

Second, we have observed that disruptive events can be triggered by processes running on the virtual machine or by GitLab CE's own management, which can have a significant impact on the performance of the server.

Third, we have noted significant changes in the performance of the server at the start of the evaluation, where there is a sudden change observed after just a few minutes or hours. These changes may be indicative of major state changes that occur during the evolution of community interaction on the server.

\subsubsection{Found Limitations and Unknown Errors}

\begin{table*}\vspace{2mm}
\caption{Fuzzing results. The 6,907 errors are the result of executing 1,696,386 events. 
}
\label{tab:fuzzing}\vspace{-2mm}
\begin{tabular}{lp{60mm}rp{65mm}}
\toprule
API return value    & API explanation~\cite{gitlabapicodes}      & count          & Our best explanation                                                     \\ \midrule
304 Not   Modified & The resource hasn’t been modified since the last request.                                                                                      & 1     (0.01\%)  & 
We have identified the offending user and their actions, but we have no explanation. %
\\
400 Bad   Request  & A required attribute of the API request is missing. For   example, the title of an issue is not given.                                         & 3  (0.04\%)  & It is unclear why the three ``issues'' (PushEvents) fail, as we are enforcing minimum and maximum length on issue title, body, and message. \\
403   Forbidden    & The request isn’t allowed. For example, the user isn’t allowed   to delete a project.                                                          & 172  (2.49\%)  & Because of a known issue: GitLab lag~\cite{gitlablag}.                                                                                                                  \\
404 Not   Found    & A resource couldn’t be accessed. For example, an ID for a   resource couldn’t be found, or the user isn’t authorized to access the   resource. & 3660 (52.99\%) & Because of a known issue: GitLab lag~\cite{gitlablag}.                                                                                                                   \\
409   Conflict     & A conflicting resource   already exists. For example, creating a project with a name that already   exists.                                    & 297  (4.30\%)  & 
Because of a known issue: GitLab lag~\cite{gitlablag}.                                                                                                                   \\
NaN                & \textit{(not documented)}                                                                                                                                              & 2774  (40.16\%) & We have no explanation.                      \\ \bottomrule                                                      
\end{tabular}\vspace{-2mm}
\end{table*}

\paragraph{Found Limitation} 
Preliminary testing throughout our study has uncovered an issue with the server that could not have been detected through a simple repetition of the Simple dataset: GitLab imposes limits to ensure optimal performance quality. Our initial experiments revealed a restriction on the maximum number of followed users to 300. This was confirmed by a review of public issue comments in GitLab, which indicated that this limit was in place to prevent the activity page for followed users from failing to load~\cite{Issue360755}. In our case, tests resulted in HTTP 304 errors when trying to load follow events for users who were following more than 300 users, causing the event creation to fail. 
To resolve this issue, we manually edited GitLab's source code to increase the limit to 1,523, the total number of users in our dataset. 
We see this as a good illustration of how diverse data can outperform original data for testing configurations: a broader range of data points offers more chances of uncovering issues, making it a valuable asset for any testing process.

\paragraph{Fuzzing Results} 
During the loading of the interaction graphs, which involved the execution of 1,696,386 events, we observed 6,907 (0.41\%) errors (see Table~\ref{tab:fuzzing}). 
First, we attribute the majority of errors (59.78\%) to known issue: GitLab's internal processing results in a lag~\cite{gitlablag}; for example, the creation of a repository may be successful, but a future PushEvent may fail as it is not available yet. 
Practically, we could counter this by adding a delay into our framework, but a better solution would be to improve the performance on the server's side. 
Second, there are four outstanding errors for which we have identified the sources but we cannot offer any explanations despite our best efforts. 
Third, a substantial portion of errors (40.16\%) escapes our debugging attempts: the API responds with empty dictionaries, which appears to be (for our use cases) undocumented behaviour.

\section{Efficient evolution and evaluation}

In conclusion of this first case study of Socialz, we would like to bring attention to two important research questions: how can we efficiently (1) evolve and (2) evaluate community interaction?

Efficiently evolving community interaction is crucial as it allows for iterative improvements and explorations based on observed data, increasing the chances of identifying social bugs. 
It may be beneficial to evolve interactions either 
in-vivo, i.e. small-scale interactions are evaluated ``live'' on a running server, 
or to run comprehensive evaluations for each large community interaction akin to those performed in Section~\ref{sec:eval}. 
However, small-scale evaluations come with the challenge of affecting the virtual machine and creating unintended side-effects, such as triggering memory clean-ups or altering the system's performance over time, as seen in Figure~\ref{fig:prometheus}. 
In contrast to this, comprehensive analyses of the entire community interaction (which may involve hundreds of thousands of events like in our case) are time-consuming, taking over one day each. 
As a middle way between these two extremes, differential evaluations may be a solution, but only if the effects of mutations can be attributed efficiently and accurately. Currently, this presents a significant challenge both in practice and algorithmically.

The question of how much to evolve community interaction is closely tied to the question of how to evaluate  community interaction. As our data analysis has shown, there is a significant amount of noise present in the server under examination. This is a common issue in complex systems such as Android phones~\cite{bokhari2020validation}, where the targeted application shares resources with multiple processes and modern multi-core hardware. Existing validation methods, such as complete rollbacks to known states or extended repetition, are not practical due to the time and resources required. This necessitates schemes and performance indicators that allow for reliable and efficient attribution of community interactions to their effects.

\section{Threats to Validity}

In this section, we discuss potential threats that might have affected the validity of our work.

\subsection{External Validity}

External validity concerns the extent to which the results of a study can be generalised or applied to settings other than those used in the study. Our work was conducted on a specific dataset from GitHub and a GitLab CE server. This specific environment and dataset might not be representative of all possible scenarios. However, our choice was based on practical considerations (for example, we are unaware of another, substantial social network for which interaction data is available as well as options to execute those interactions on a production-quality server) and the widespread usage of these platforms in the developer community. In addition, the differences among the Original, Evolved, Simple, and Random datasets might not capture the entire spectrum of possible community interactions. This limitation might affect the generalisability of our findings across other datasets.

\subsection{Internal Validity}

Internal validity pertains to whether a study establishes a cause-and-effect relationship. The creation of the ``Random'' dataset by generating connections randomly could introduce uncontrollable factors that might not be representative of real-world interactions. While it serves as a basis of comparison, it is crucial to be cautious when interpreting results derived from this dataset. To overcome a limitation in the number of followed users, we modified GitLab's source code. Although we conjecture that this alteration did not adversely affect our results, it is essential to acknowledge that this change could introduce unforeseen variables into the experiment.

\subsection{Construct Validity}

Construct validity evaluates whether the measures used for the study accurately represent the concept they are intended to measure. The method of crafting the ``Simple'' dataset by merely copying events until it matches the size of the evolved list might not be the most robust way of ensuring simplicity.

\section{Conclusions and Future Work}

We present a new social fuzz testing method called Socialz, which is based on publicly accessible data from GitHub. Our approach uses evolutionary computation to diversify community interactions, and then evaluates the results on a GitLab~CE server.

The key takeaways of our research are: 
\begin{enumerate}[topsep=2mm]
    \item Although the initial setup of social fuzz testing requires significant effort, it is feasible.
    \item Evolutionary diversity optimisation can generate community interactions that are significantly different from the original data or random data, potentially uncovering social bugs. 
    \item In the actual fuzzing, we came across (1) one known limitation of GitLab~CE (that simple data replay could not), and (2) 6,907 errors, of which 40.16\% are beyond our debugging skills. 
\end{enumerate}

Future work in this area offers endless possibilities, such as the further characterisation of sub-communities, the exploration of additional community interactions and the related features, and the integration of Socialz with traditional fuzz testing techniques that target code-level or system-level interactions.

To support future research in social testing, we have made available all code, data, and virtual machines used in this study: \url{https://github.com/fzanart/Socialz/}.

\vspace{2mm}\noindent\textbf{Acknowledgements: }
This project has been enabled by a gift from Facebook/Meta: \url{https://research.facebook.com/blog/2021/9/announcing-the-winners-of-the-2021-rfp-on-agent-based-user-interaction-simulation-to-find-and-fix-integrity-and-privacy-issues/}

\bibliographystyle{ACM-Reference-Format}
\bibliography{sample-base}


\end{document}